\documentclass[a4paper]{jpconf}
\usepackage{graphicx}
\usepackage[utf8]{inputenc}
\usepackage[square,sort&compress,comma,numbers]{natbib}
\usepackage{amsmath}
\usepackage{amssymb}
\usepackage[dvipsnames]{xcolor}
\begin{document}
\title{Exploring continuum structures in reactions with three-body nuclei}

\author{J. Casal$^1$, M. Gómez-Ramos$^{2,3}$, A. M. Moro$^3$, A. Corsi$^4$}

\address{$^1$ Dipartimento di Fisica e Astronomia ``G.~Galilei'', Università degli Studi di Padova and INFN Sezione di Padova, via Marzolo 8, I-35131 Padova, Italy}

\address{$^2$ Institut  für  Kernphysik,  Technische  Universität  Darmstadt,  D-64289 Darmstadt, Germany}

\address{$^3$ Departamento de Física Atómica, Molecular y Nuclear, Facultad de Física, Universidad de Sevilla, Apdo. 1065, E-41080 Sevilla, Spain}


\address{$^4$ Département de Physique Nucléaire, IRFU, CEA, Université di Paris-Saclay, F-91191, Gif-sur-Yvette, France}

\ead{casal@pd.infn.it}

\begin{abstract}
The Transfer to the Continuum method has been applied to describe the $^{11}\text{Li}(p,pn)$ and $^{14}\text{Be}(p,pn)$ reactions in inverse kinematics, using structure overlaps computed within a full three-body model for the projectile. Calculations agree with the available experimental data on the unbound $^{10}$Li and $^{13}$Be nuclei.
\end{abstract}

\section{Introduction}
Nuclear systems at the limit of stability display exotic properties which have motivated extensive theoretical and experimental developments. Light nuclei lying on and beyond the driplines offer a unique enviroment to study nucleon-nucleon correlations and clustering. Among them, the properties of Borromean two-neutron halo nuclei such as $^6$He, $^{11}$Li or $^{14}$Be have attracted special interest~\cite{tanihata13}. In these three-body systems, the loosely bound valence neutrons explore distances far from a compact core, giving rise to a diffuse matter distribution with a strong effect on interaction cross sections and electromagnetic responses due to the coupling to the continuum. The properties of the unbound $\text{core}+n$ nuclei shape the structure of two-neutron halos, with the correlation between the valence neutrons playing a key role in binding the system~\cite{Zhukov93}. 

In this contribution, we present some recent development for the theoretical description of nucleon-removal reactions populating unbound nuclei. First, we briefly recall the Transfer to the Continuum (TC) framework to study $(p,pn)$ knockout reactions from two-neutron halo nuclei in inverse kinematics. 
Then, we consider the cases of $^{11}\text{Li}(p,pn){^{10}}\text{Li}$ and $^{14}\text{Be}(p,pn){^{13}}\text{Be}$. 

\section{Transfer to the Continuum (TC)}
\begin{figure}[t]
	\centering
	\includegraphics[width=0.4\linewidth]{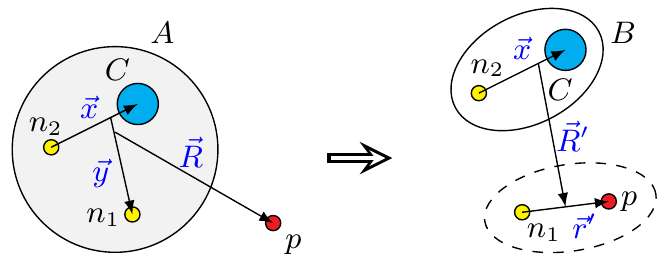}\hspace{2pc}%
	\begin{minipage}[b]{14pc}\caption{\label{fig:scheme}Diagram for a $(p,pn)$ reaction induced by a $\text{core}+n+n$ projectile in inverse kinematics.}
	\end{minipage}
\end{figure}

The $\text{core}+n$ relative-energy spectra and momentum distributions in $(p,pn)$ knockout reactions can be computed within the Transfer to the Continuum (TC) framework~\cite{AMoro15}, which was recently extended to the case of three-body projectiles~\cite{gomezramosplb17}. The process takes the form
\begin{equation}
\underbrace{(C+n_1+n_2)}_A+p ~\longrightarrow~ \underbrace{(C+n_2)}_B+n_1+p
\end{equation}
and is depicted in Fig.~\ref{fig:scheme}, where the residual nucleus ($B$) is unbound. The transition amplitude for such a process can be written in prior form as
\begin{equation}
\mathcal{T}_{if}=\left\langle \varphi_{B,\boldsymbol{q}}^{(-)}(\boldsymbol{x}) \Upsilon_{f}^{(-)}(\boldsymbol{r}',\boldsymbol{R}')\Big|V_{pN_1}+U_{pB}-U_{pA} \Big|\Phi_A^{3b}(\boldsymbol{x},\boldsymbol{y}) \chi_{pA}^{(+)}(\boldsymbol{R}) \right\rangle,
\label{eq:TA}
\end{equation}
where $\Phi_A^{3b}$ represents the ground-state wave function of the three-body projectile, $\chi_{pA}^{(+)}$ is the distorted $p$-$A$ wave in the entrance channel, $\varphi_{B,\boldsymbol{q}}^{(-)}$ is the two-body continuum wave function of the binary subsystem with momentum $\boldsymbol{q}$, and $\Upsilon_{f}^{(-)}$ is a scattering wave function describing the motion in the final $p+n_1+B$ partition. The latter is expanded in $p$-$n_1$ continuum states using the binning procedure typically adopted in continuum-discretized coupled-channels (CDCC) calculations. More details on the formalism can be found in Ref.~\cite{gomezramosplb17}.

Then, considering that the $(p,pn)$ process is driven by the interaction of the proton target with a single valence nucleon of the projectile, which implies a participant/spectator approximation, the potentials appearing in Eq~(\ref{eq:TA}) do not change the state of $B$, and one can perform the integral with respect to $\boldsymbol{x}$. This provides a set of overlap functions between the initial projectile wave function and the continuum states of the $\text{core}+n$ residue,
\begin{equation}
\psi_{LJJ_T}^{q} (\boldsymbol{y}) = \langle \varphi_{B,q}^{LJJ_T}|\Phi_A^{3b} \rangle
\label{eq:overlaps}
\end{equation}
where the single-particle configurations $\{L,J\}$ are coupled with the spin of the core to give the total angular momentum $J_T$ of $B$. With this definition, the total transition amplitude can be expanded in different $LJJ_T$ components,
\begin{equation}
\mathcal{T}_{if}^{LJJ_T}=\left\langle \Upsilon_{f}^{(-)}\Big|V_{pN_1}+U_{pB}-U_{pA} \Big|\psi_{LJJ_T}^{q} \chi_{pA}^{(+)} \right\rangle,
\label{eq:TAccba}
\end{equation}
where all the structure information is contained in the overlaps and can be incorporated in a consistent way by using the same $\text{core}+n$ interaction in both the two- and three-body calculations for $B$ and $A$, respectively.

\section{Application to $\boldsymbol{^{10}}$Li and $\boldsymbol{^{13}}$Be}
The properties of the unbound $^{10}$Li and $^{13}$Be systems can be explored in $(p,pn)$ reaction from the two-neutron halo nuclei $^{11}$Li and $^{14}$Be~\cite{aksyutina13a}. We recently peformed calculations and compared with the available experimental data for the $^{11}\text{Li}(p,pn){^{10}}\text{Li}$~\cite{gomezramosplb17} and $^{14}\text{Be}(p,pn){^{13}}\text{Be}$~\cite{corsi19} reactions measured at GSI and RIKEN facilities, respectively. In these works, we explored the sensitivity of the relative-energy cross sections and momentum distributions on the structure properties of the initial and final nuclei. Here we present a summary of the most relevant results with the adopted models for $^{10}$Li and $^{13}$Be. In both cases, the ground-state wave function for the $\text{core}+n+n$ projectile was obtained by solving the three-body problem within the hyperspherical framework using a pseudostate basis (see, for instance, Refs.~\cite{Zhukov93,JCasal13}). 

\subsection{${^{10}}$Li}
\begin{figure}[t]
	\centering
	\includegraphics[width=0.5\linewidth]{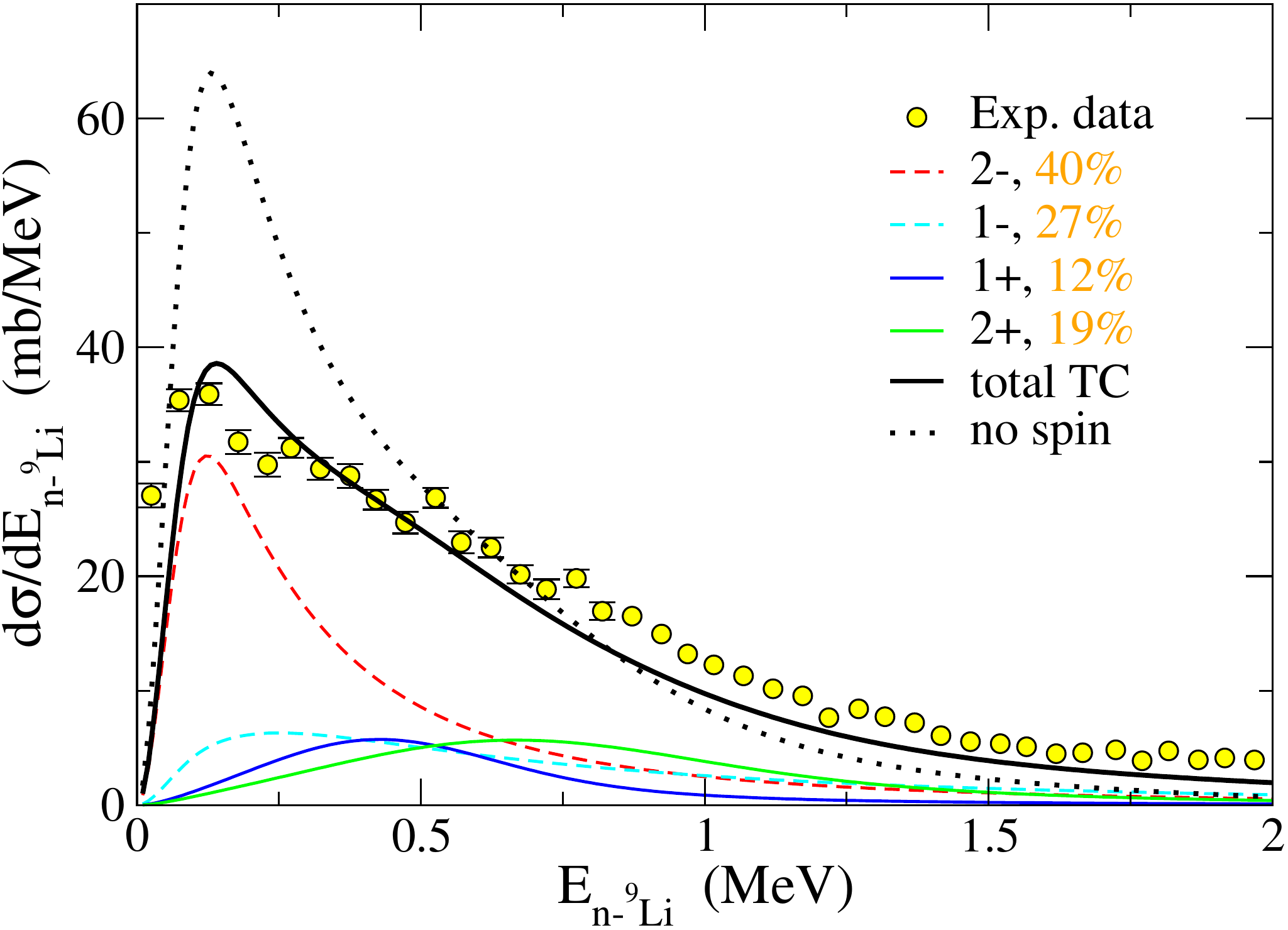}\hspace{2pc}%
	\begin{minipage}[b]{14pc}\caption{\label{fig:10li}Relative $n$-$^9$Li energy spectrum populated via $^{11}\text{Li}(p,pn){^{10}}\text{Li}$ at 280 MeV/u. Solid (dashed) lines correspond to $p_{1/2}$ ($s_{1/2}$) contributions, while the thick black line is the total cross section. The dotted line is a calculation ignoring the spin of the core, and the experimental data are from Ref.~\cite{aksyutina2008}.}
	\end{minipage}
\end{figure}

Our results for $^{11}\text{Li}(p,pn){^{10}}\text{Li}$ are presented in Fig.~\ref{fig:10li}, compared with the experimental data measured at GSI at 280 MeV/u~\cite{aksyutina2008}. Our three-body model includes explicitly the spin of the $^9$Li core, $I=3/2^-$, so that the $n$-$^9$Li single particle configurations $s_{1/2}$ and $p_{1/2}$ split in $1^-,2^-$ and $1^+,2^+$ doublets, respectively. In this model, the ground state of $^{10}$Li is a 2$^-$ virtual state, displaying the parity inversion reported also for the $N=7$ isotone $^{11}$Be. In addition, the model produces two overlapping $p$-wave resonances around 0.5 MeV above the neutron separation threshold. These features produce a 3/2$^-$ ground-state of $^{11}$Li with 31\% of $p_{1/2}$ components. The computed TC relative-energy spectrum for $^{10}$Li, after the convolution with the experimental resolution, agrees reasonably well with the experimental data. Note that the TC calculations provide absolute cross sections, so no fitting procedure is required to explain the data. As shown in Ref.~\cite{gomezramosplb17}, the comparison with other models for $^{10}$Li supports the persistence of the parity inversion beyond the neutron dripline, which is also key in describing the available $(p,d)$ and $(d,p)$ transfer data~\cite{casalplb17,moroplb19}. The dotted line in Fig.~\ref{fig:scheme} corresponds to a calculation in which the spin of the core is ignored, a common assumption in several experimental and theoretical works. In that case, the shape of the relative-energy spectrum cannot be reproduced.

\subsection{${^{13}}$Be}
In the case of the $^{14}\text{Be}(p,pn){^{13}}\text{Be}$ reaction, experimental results have shown contradictory conclusions due to the long-debated structure of the exotic $^{13}$Be system~\cite{kon10,aksyutina13b,Ribeiro18}. The low-lying peak observed in $^{12}\text{Be}+n$ relative energy spectra has been interpreted either as a 1/2$^-$ ($\ell=1$) resonance caused by the quenching of the shell gap for $N=8$~\cite{kon10} or a 1/2$^+$ ($\ell=0$) state~\cite{aksyutina13b}. The detection of gamma rays coming from the decay of $^{12}$Be, pointing toward core-excited components in $^{13,14}$Be, adds complexity to the description of their structure and reaction dynamics and to the interpretation of the experimental data. To reduce the ambiguities in the usual analysis of $(p,pn)$ reactions based on standard fitting procedures, we studied recent $^{14}\text{Be}(p,pn)$ data within the TC approach. Details on the experimental setup and data analysis can be found in Ref.~\cite{corsi19}, including the discussion about the gamma decay of $^{12}\text{Be}(2^+,1^-)$, which was measured with high statistics.

\begin{figure}[t]
	\centering
	\includegraphics[width=0.5\linewidth]{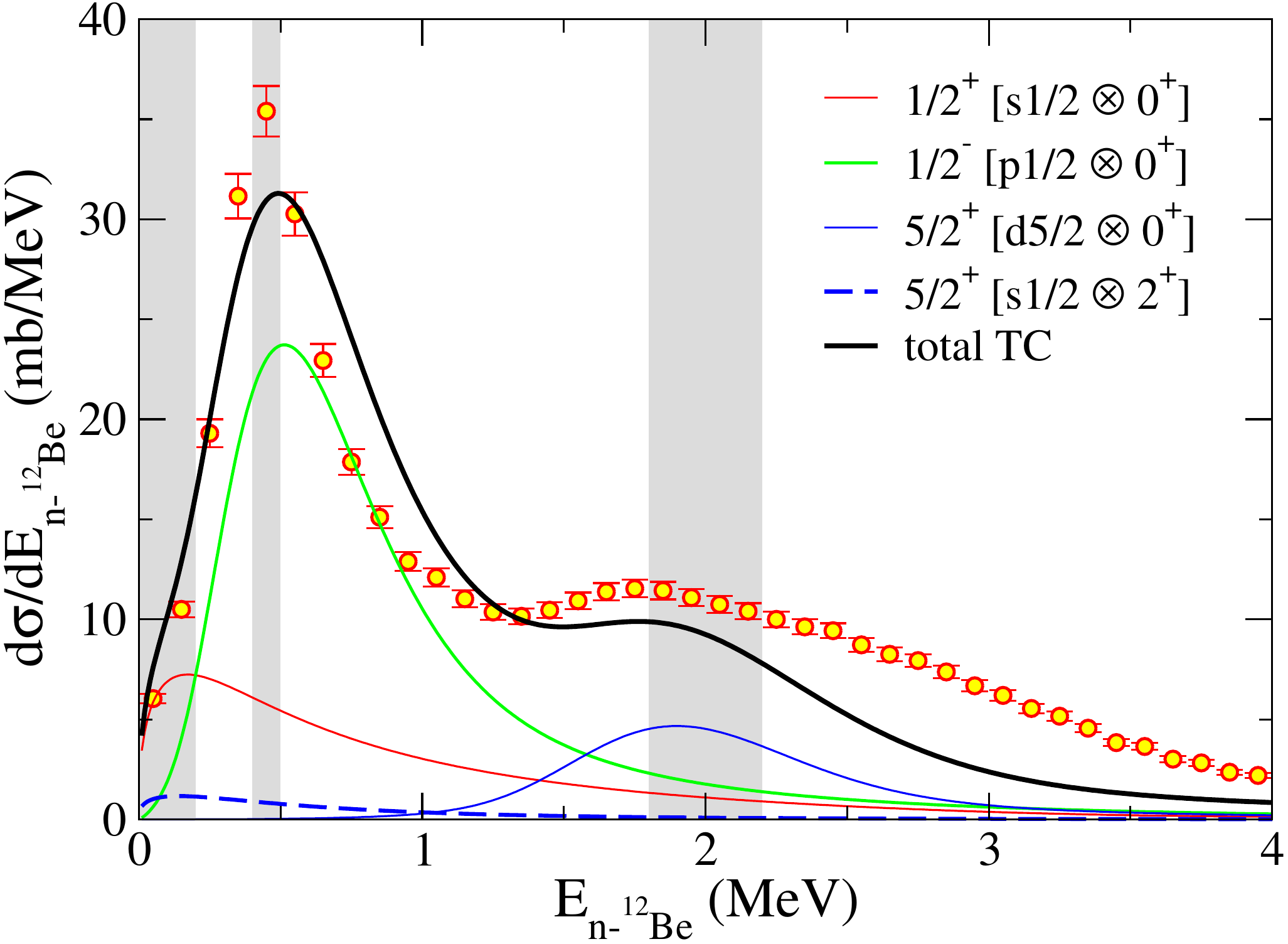}\hspace{2pc}%
	\begin{minipage}[b]{14pc}\caption{\label{fig:13be}Relative $n$-$^{12}$Be energy spectrum populated via $^{14}\text{Be}(p,pn){^{13}}\text{Be}$ at 265 MeV/u. The four leading terms are shown, corresponding to different total angular momenta of the $^{13}$Be system, together with the total cross section. Shaded areas are the energy regions considered for the momentum distributions in Fig.~\ref{fig:mom}.}
	\end{minipage}
\end{figure}

In order to include some core excitations in the model, we introduced an effective quadrupole deformation in the $\text{core}+n$ potential following the prescription of Ref.~\cite{IJThompson04}, allowing couplings to the first $2^+$ excited state in $^{12}$Be. The adopted model for $^{13}$Be, which was found to give the best agreement with the $(p,pn)$ data, is dominated by a low-lying $p$-wave state at $\sim$ 0.5 MeV, together with the well-established $5/2^+$ ($\ell=2$) resonance around 2 MeV. As shown in Ref.~\cite{corsi19}, the resulting $^{14}$Be ground-state wave function contains a large admixture of $p$ and $sd$ components, and the core-excited contribution amounts for $\sim$ 20\% of the norm. Using this model, the TC relative-energy spectrum is well reproduced at low excitation energies, as shown in Fig.~\ref{fig:13be}. Note that several terms, arising from the coupling of the spin of the core with single-particle components, contribute to the total cross section, but for clarity only the most important ones are presented. The underestimation of the strength above 2 MeV relative energy is likely to come from missing components in the wave function, in particular those coupled with other excited states of the core. 

\begin{figure}[ht]
	\centering
	\includegraphics[width=0.75\linewidth]{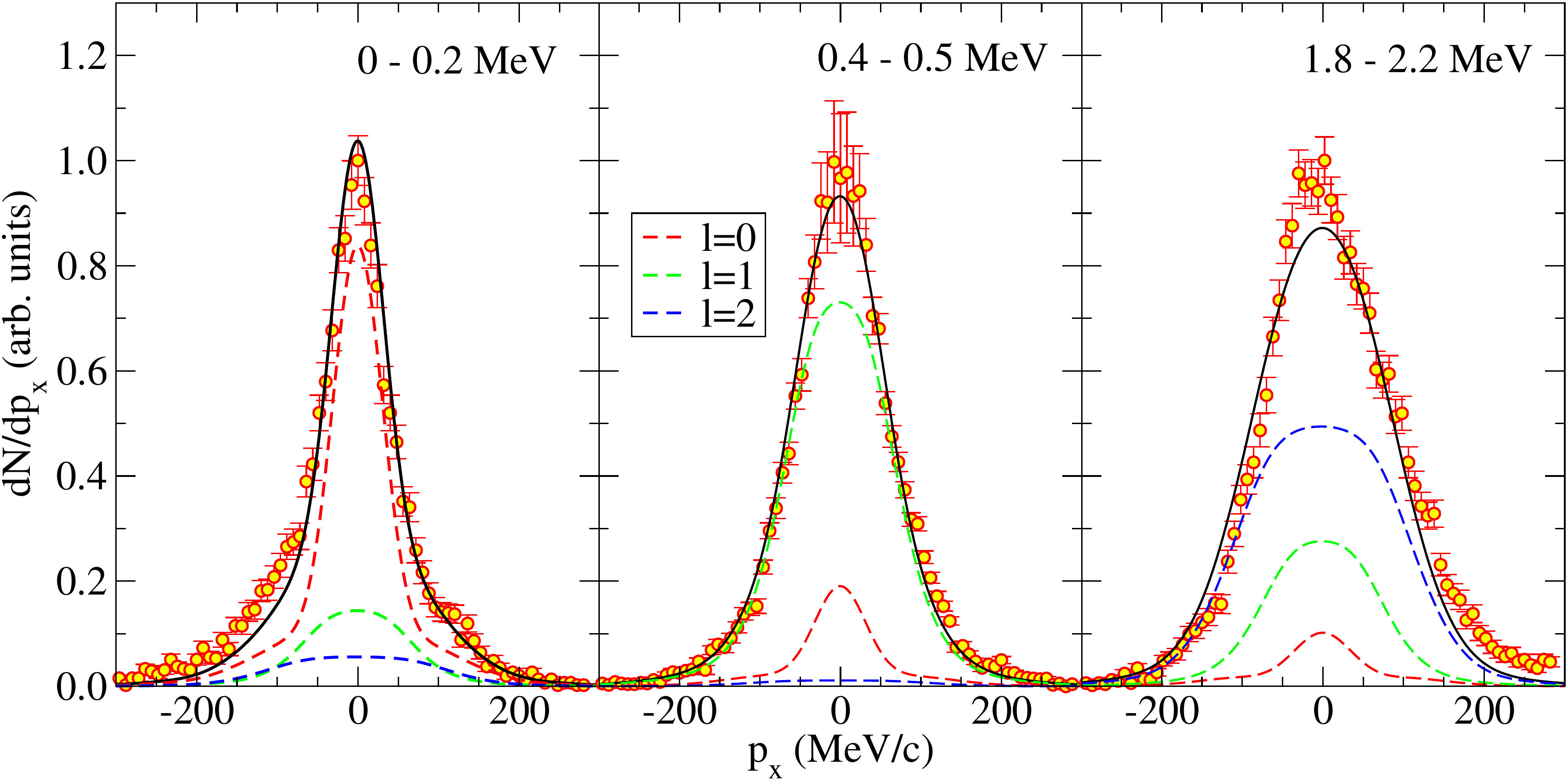}
	\caption{\label{fig:mom} Transverse momentum distribution of the knocked-out neutron in $^{14}\text{Be}(p,pn){^{13}}\text{Be}$, corresponding to relative $n$-${^{12}}$Be energies of 0-0.2, 0.4-0.5 and 1.8-2.2 MeV (shaded areas in Fig.~\ref{fig:13be}). The dashed lines correspond to removal of a neutron from $s$, $p$ or $d$ wave, while the solid line is the total TC result rescaled through a $\chi^2$ fit.} 
\end{figure}

The most prominent feature of our $^{13}$Be relative-energy spectrum is the good description of the low-lying peak with a $p$-wave resonance. The validity of such assumption in the theoretical model can be tested by studying the momentum distributions of the knocked-out neutron. This is presented in Fig.~\ref{fig:mom} for three relative-energy bins. Our calculations include the added $\ell=0,1,2$ contributions, with their relative weights given by the cross sections in Fig.~\ref{fig:13be}, and the total momentum distribution rescaled to the data through a $\chi^2$ fit. It is remarkable that the width of the momentum distributions are overall well reproduced. While the 0 - 0.2 and 1.8 - 2.2 MeV regions are well described by (mostly) $s$- or $d$-wave contribution, we find that the data around the 0.4 - 0.5 peak is consistent with $p$-waves. 

This result contrasts with the findings in Ref.~\cite{aksyutina13b}, where an $s$-wave resonance was proposed to explain the width of the momentum distribution in the energy region 0.4 - 0.5 MeV. This apparent inconsistency may arise from the different methods used to interpret the data. The analysis presented in Ref.~\cite{aksyutina13b} relays on a $\chi^2$ procedure to assign the weight of $s$-, $p$- or $d$-components, while in our calculations the weights of the different components are fixed by the structure model and the reaction calculations. The limitations of the fitting procedure are already recognized in Ref.~\cite{aksyutina13b}, in particular they acknowledge the ``large statistical uncertainties that prevent a strict conclusion'' due to the correlation of the fitting parameters. Furthermore, the analysis in Ref.~\cite{aksyutina13b} is based on the theoretical method by Hansen~\cite{hansen96}, which was devised for knockout reactions on heavier targets and yields momentum distributions for $\ell=0,1,2$ with different widths than those obtined in our TC calcultations. It is our understanding that the present approach avoids many of the ambiguities in the previous work and reinforces the necessity for prior models of the three-body projectile in the analysis of this kind of experiments.

\section{Conclusions}

We have presented Transfer to the Continuum (TC) calculations to describe $(p,pn)$ reactions induced by two-neutron halo nuclei, in which $\text{core}+n$ unbound states are populated. Our method is based on a participant/spectator approximation of the transition amplitude, with all the structure information contained in the overlaps between the ground-state of the initial three-body nucleus and the continuum states of the two-body residual system. Within this framework, the usual ambiguities involved in the analysis of $\text{core}+n$ relative-energy spectra using fitting procedures are avoided. Our results for $^{11}\text{Li}(p,pn){^{10}}\text{Li}$ describe very well the available experimental data and confirm the parity inversion for $N=7$ beyond the neutron dripline. For $^{14}\text{Be}(p,pn){^{13}}\text{Be}$, our analysis of the relative-energy spectrum and the corresponding momentum distributions is consistent with a dominance of $p$-waves at low excitation energies.

\section*{Acknowledgements}

This work has been partially supported by the Spanish Ministerio de Ciencia, Innovaci\'on y Universidades  and the European Regional Development Fund (FEDER) under Projects  No.~FIS2017-88410-P and FPA2016-77689-C2-1-R, and by the European Union's Horizon 2020 research and innovation program under grant agreement No.~654002.
\vspace{-5pt}

\bibliographystyle{iopart-num}
\bibliography{bibfile}

\end{document}